\documentclass[10pt]{iopart}

\usepackage{graphicx}
\usepackage{float}
\usepackage{amssymb}
\usepackage{cite}
\usepackage{color}
\usepackage[dvipsnames]{xcolor}
\newcommand{\AI}[1]{{ #1}}

\begin{document}

\title[{Scaling Properties of Polycrystalline Graphene: A Review}]{{Scaling Properties of Polycrystalline Graphene: A Review}}
\author{ Andreas Isacsson$^1$, Aron W. Cummings$^2$, Luciano Colombo$^{2,3,4}$, Luigi Colombo$^5$, Jari M. Kinaret$^1$, and Stephan Roche$^{2,6}$}
\address{$^1$Department of Physics, Chalmers University of Technology, SE-412 96 Gothenburg, Sweden}
\address{$^2$Catalan Institute of Nanoscience and Nanotechnology (ICN2), CSIC and the Barcelona Institute of Science and Technology Campus UAB, 08193 Bellaterra, Barcelona, Spain}
\address{$^3$Dipartimento di Fisica, Universit\`a di Cagliari, Cittadella Universitaria, 09042 Monserrato (Ca), Italy}
\address{$^4$Institut de Ci\`encia de Materials de Barcelona (ICMAB--CSIC), 08193 Bellaterra, Barcelona, Spain}
\address{$^5$Texas Instruments Incorporated, 13121 TI Boulevard, MS-365, Dallas, Texas 75243, United States}
\address{$^6$ICREA, Instituci{\'{o}} Catalana de Recerca i Estudis Avan\c{c}ats, 08070 Barcelona, Spain}
\ead{andreas.isacsson@chalmers.se}

\begin{indented}
\item[\today]
\end{indented}

\begin{abstract}
We present an overview of the electrical, mechanical, and thermal
properties of polycrystalline graphene. Most global properties of this
material, such as the charge mobility, thermal conductivity, or
Young's modulus, are sensitive to its microstructure, for instance the
grain size and the presence of line or point defects. Both the local
and global features of polycrystalline graphene have been investigated
by a variety of simulations and experimental measurements. In this
review, we summarize the properties of polycrystalline graphene, and
by establishing a perspective on how the microstructure impacts its
large-scale physical properties, we aim to provide guidance for
further optimization and improvement of applications based on this
material, such as flexible and wearable electronics, and
high-frequency or spintronic devices.
\end{abstract}

\submitto{\TDM}

\maketitle
\ioptwocol
\section{Introduction}
The macroscopic physical properties of polycrystalline materials
depend crucially on the structure and distribution of crystallites and
grain boundaries. These, in turn, depend on the synthesis method
used. To produce high-quality and large-scale two-dimensional
materials such as graphene, transition metal dichalcogenides, and
hexagonal boron nitride, the manufacturing method of choice has
primarily been chemical vapor deposition (CVD). CVD-grown graphene and
related materials exhibit a polycrystalline morphology, consisting of
a patchwork of individual grains which coalesce to form
one-dimensional boundaries separating domains of different crystalline
orientations. Although graphene crystal growth has been at the center
of much research over the past decade, large flat perfect single
crystals remain fairly elusive, and the resulting structures depend on
the details of the fabrication process. Large-area graphene grown by
CVD on metals such as Cu, Cu-Ni, Pt, Ru, and Ir, is typically
polycrystalline with grain sizes ranging from a few hundred nanometers
to several centimeters in diameter, with a roughness that mimics that
of the metal substrate. Graphene films grown on SiC, on the other
hand, have shown excellent flatness but replicate the crystallographic
step structure of the SiC single crystals, which leads to few-layer
graphene or edge defects.

For understanding macroscopic properties, studying individual
boundaries is thus not sufficient, and one must also take a more
global view of the effects of the polycrystalline structure,
e.g. grain size distribution. In this review, we will therefore first
present the essentials of grain boundary geometries, which are
characterized principally by non-hexagonal rings such as pentagons and
heptagons. Afterwards, the average grain size will be taken as a
reference parameter to analyse the scaling of charge transport,
mechanical properties, and thermal conduction as a function of
geometry, contrasting when possible the theoretical predictions with
available experimental data.

\begin{figure*}[t]
\begin{center}
\includegraphics[width=\linewidth]{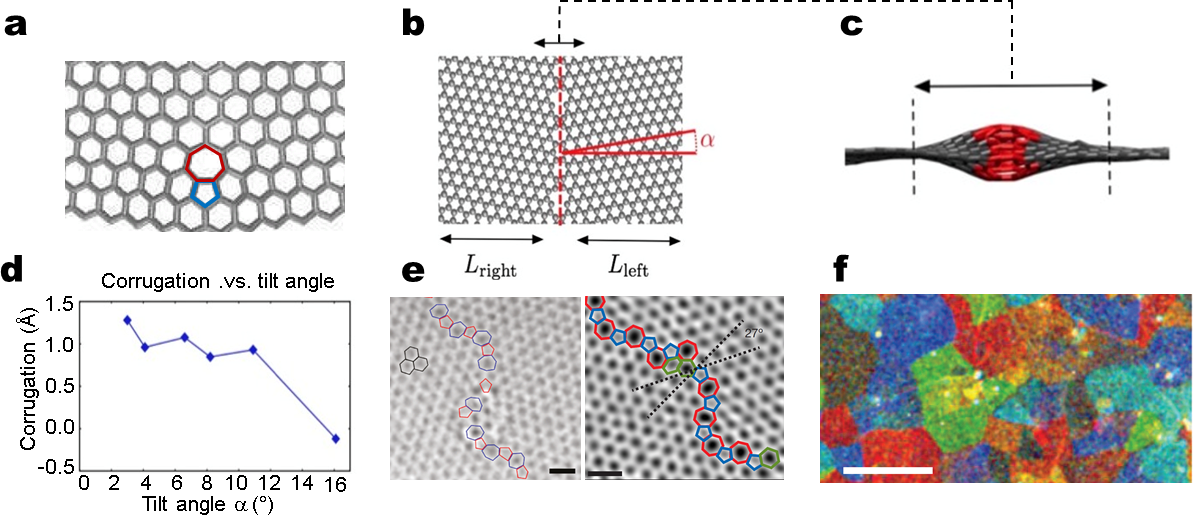}
\caption{ { Structure of grain boundaries in polycrystalline
    graphene.} {\bf a.} A disclination in graphene consists of a
  pentagon-heptagon pair, which maintains the three-fold
  coordination. {\bf b.} In-plane structure of a symmetric model grain
  boundary in graphene, obtained by tilting two sheets of equal length
  by an angle $\alpha=8.2^{\circ}$. {\bf c.} Out-of-plane structure
  showing corrugation occurring nearby the grain boundary, as
  predicted by MD simulations. In low-angle symmetric tilt grain
  boundaries, out-of-plane buckling (warping) reduces the energy
  associated with boundary stress. {\bf d.} Corrugation as function of
  the tilt angle. {\bf e.} Meandering grain boundaries imaged by
  AC-TEM.  In CVD-grown graphene, the boundaries are typically neither
  symmetric nor straight, but have more complex geometry \AI{[adapted
      with permissions from ACS Nano {\bf 5} 2142, copyright (2011)
      and from Macmillan Publishers Ltd: Nature {\bf 469}, 389,
      copyright (2011)]}. {\bf f.} False-color image of
  polycrystalline graphene . The global properties of polycrystalline
  graphene are determined not only by the microscopic properties of
  the grain boundaries, but also by the distribution of grain sizes
  and crystal orientations \AI{[adapted from Science {\bf 340}, 1073,
      copyright (2013). Reprinted with permission from AAAS]}.}
\label{fig:GB}\end{center}
\end{figure*}

\section{Grain boundaries in graphene\label{sec:GBconf}}
Considerable effort has so far been spent on understanding the
structure and energetics of single grain boundaries (GBs) separating
two grains. Macroscopic characterization of such a boundary requires
knowledge of the misorientation angle $\alpha$ between the two regions
(left and right) separated by the grain boundary, the direction $\psi$
of the GB, and a translation vector that gives the relative
displacement between the lattices on different sides of the
boundary~\cite{Gottstein2004}.  In the simplest case, the GB is
symmetric and forms a mirror symmetry plane between the two
crystalline regions. This requires that the crystal orientations to
the left and right of the grain boundary are $\alpha_{\rm L} = \psi -
\alpha$ and $\alpha_{\rm R} = \psi + \alpha$ \footnote{Some references
  use another definition of crystalline axes in the two regions, which
  results in $\alpha_{\rm R} \rightarrow -\alpha_R$ and $\alpha =
  (\alpha_{\rm L}+\alpha_{\rm R})$ rather than our convention where
  $\alpha = (\alpha_{\rm R} - \alpha_{\rm L})/2$}. A symmetric grain
boundary exhibits periodicity with period $(n^2 + nm + m^2)^{1/2}$
where $n$ and $m$ give the direction of the boundary in terms of the
basis vectors of the left or right region.  For small misorientation
angles $\alpha$, a symmetric grain boundary can be thought of as
arising from a periodic series of parallel edge dislocations
terminating at the boundary.

Microscopically, grain boundaries in graphene result in deviations
from the regular hexagonal structure of the graphene lattice. \AI{In
  crystalline membranes, an individual positive or negative
  disclination cannot be viewed as a point defect as it results in a
  global warping with a logarithmically diverging
  energy~\cite{Nelson1988}. Hence, pairs of positive and negative
  disclinations are needed~\cite{Nelson1988}. In graphene, edge
  dislocations are thus typically terminated by pentagon-heptagon
  pairs~\cite{Liu2010,Yazyev2010_PRB,Yazyev2014}. These disclination
  dipoles conserve the three-fold coordination of each carbon atom and
  yield a low energy cost for the defect (see Fig.~\ref{fig:GB}a)}.

A small misorientation angle $\alpha$ results in widely separated
pentagon-heptagon pairs, while a large $\alpha$ requires the pairs to
be close to each other (see Fig.~\ref{fig:GB}b). As shown in
\cite{Pao2011, Yazyev2010}, in order to match periodic boundary
conditions (needed for modeling by atomistic simulations) only some
specific orientations are possible, corresponding to $\alpha =
3.0^{\circ}, 4.1^{\circ}, 8.2^{\circ}, 10.9^{\circ}$, and
$16.1^{\circ}$. However, for a generic large $\alpha$ the boundary
must assume a more complicated structure than simple pentagon-heptagon
pairs. In the general case of an asymmetric grain boundary,
periodicity only occurs if certain commensurability conditions are
met, and the resulting period is typically much longer, which usually
results in a higher energy cost per length of the boundary and a more
complicated analysis~\cite{Malola2010, Yazyev2010}.

To accommodate the strain generated by the GB, out-of-plane
corrugation of the graphene sheet typically
occurs\cite{Nelson1988,Liu2010,Yazyev2010_PRB}.  Figure~\ref{fig:GB}c
shows C-atom displacements perpendicular to the graphene
plane. \AI{The buckling amplitude, decreasing with increasing tilt
  angle $\alpha$ is typically $\sim 1.0$ \AA\ (see
  Fig.~\ref{fig:GB}d), in good agreement with available data
  \cite{Liu2014,Tison2014}. The width of the buckled region,
  corresponding in turn to the GB thickness, is as large as 3-4 \AA,
  and can be either symmetrical or antisymmetrical around the
  disclination dipole~\cite{Lehtinen2013}.}

While many theoretical analyses to date have focused on grain boundary
constructions with rather short periodicities, grain boundary
characterizations using transmission electron
microscopy~\cite{Kim2011,Huang2011} have revealed more complicated
structures (see Fig.~\ref{fig:GB}e, f). Although the fundamental units
of these boundaries are pentagon-heptagon pairs, they tend to be
meandering. Recently, Rasool, Ophus and co-workers carried out
large-scale experimental characterization and molecular dynamics
simulations of GBs with lengths up to 200
nm~\cite{Rasool2014,Ophus2015}. For isolated grain boundaries, the
agreement between the molecular dynamics simulations and experimental
measurements is quite good in terms of the atomic structure and
mechanical properties~\cite{Rasool2014}. However, very little work has
been carried out on grain boundary junctions where three or more
crystalline regions come together and on the statistics of crystallite
size and orientation. These properties depend to a large extent on the
details of the fabrication process.

\section{Charge transport in polycrystalline graphene}
The first electrical transport measurements on isolated graphene were
made in single-crystal flakes on silicon dioxide substrates. The
subsequent introduction of graphene grown by CVD provided the
opportunity to measure large-area films \cite{Li2009}, but these films
have been predominantly polycrystalline in comparison to the device
size. In general, GBs in semiconductor materials are detrimental to
charge transport \cite{Matare1983}, and for this reason the
semiconductor industry favors high-quality single-crystal materials
with a low density of extended and point defects to minimize device
degradation.  Indeed, much of the success of the semiconductor
industry over the years has arisen from the ability to grow a large
volume of low-defect single crystals and thin films, out of materials
including Si, III-V, II-VI, and IV-IV compounds.

Graphene, unlike the semiconductors used for electronic devices, is a
semimetal and thus it is less clear if extended defects have a large
effect on transport properties. Therefore, because of its promise for
large-area electronic applications, a detailed understanding of the
electrical transport properties of polycrystalline graphene is
crucial. To this end, a great deal of experimental \cite{Huang2011,
  Yu2011, Jauregui2011, Tsen2012, Clark2013, Fei2013, Grosse2014,
  Ogawa2014, Yasaei2014, Cummings2014, Kochat2016} and theoretical
\cite{Yazyev2010, Kumar2012, Mark2012, Ihnatsenka2013, Liu2013,
  Vancso2013, Gargiulo2014, Vancso2014, Zhang2014, Paez2015,
  Nguyen2016, Sun2016} effort has has been devoted to studying charge
transport across individual graphene grain boundaries, and several
reviews have already discussed this topic in great detail
\cite{Cummings2014, Biro2013, Yazyev2014}. Therefore, here we briefly
summarize the main features of electrical transport across individual
graphene GBs before shifting our focus to a more global perspective of
charge transport in polycrystalline graphene.

\subsection{Electrical resistivity of individual grain boundaries}
Four-terminal measurements across individual GBs show an enhanced
electrical resistance compared to the surrounding grains, $R_{\rm
  total} = R_{\rm grains} + R_{\rm GB}$ \cite{Yu2011, Jauregui2011,
  Tsen2012, Clark2013, Grosse2014, Ogawa2014, Yasaei2014,
  Cummings2014, Kochat2016}. The origin of this enhanced resistance
has been probed with scanning tunneling spectroscopy (STS), and these
measurements have revealed that GBs tend to be n-doped compared to the
surrounding grains, such that they act as an electrical potential
barrier to charge transport \cite{Tapaszto2012,
  Koepke2013}. Magnetotransport measurements \cite{Cao2010,
  Jauregui2011, Yu2011}, numerical simulations \cite{Cummings2014b},
and STS measurements \cite{Tapaszto2012, Koepke2013} have shown that
GBs also induce weak localization, indicating that they are a source
of intervalley scattering. Additionally, temperature-dependent
measurements have shown that $R_{\rm GB}$ is independent of
temperature, providing further support for the hypothesis that
scattering at the GBs is dominated by structural defects and
impurities \cite{Jauregui2011, Yu2011, Tsen2012}.

The resistance of an individual graphene GB can be written as $R_{\rm
  GB} = \rho_{\rm GB}/W$, where $\rho_{\rm GB}$ is the GB resistivity
and $W$ is the device width (or equivalently, the GB length). The GB
resistivity thus provides an intrinsic measure of the charge transport
properties across the GB, independent of the device or measurement
technique. Beyond four-terminal electrical measurements, the
resistivity of individual GBs has been measured using a.c. electron
force microscopy (AC-EFM) \cite{Huang2011}, four-point scanning
tunneling potentiometry (STP) \cite{Clark2013}, and Joule expansion
microscopy \cite{Grosse2014}. The average $\rho_{\rm GB}$ of a series
of polycrystalline samples can also be extracted by measuring the
sheet resistance of each sample as a function of the average grain
size \cite{Vlassiouk2011, Duong2012, Yagi2013, Lee2014, Venugopal2012,
  Yang2016} and then fitting to a simple ohmic scaling law, $R_{\rm S}
= R_{\rm S}^0 + \rho_{\rm GB}/l_{\rm G}$, where $R_{\rm S}$ is the
sheet resistance of the polycrystalline sample, $R_{\rm S}^0$ is the
sheet resistance within the grains, and $l_{\rm G}$ is the average
grain size \cite{Cummings2014}.

\begin{figure}[t]
\includegraphics[width=\columnwidth]{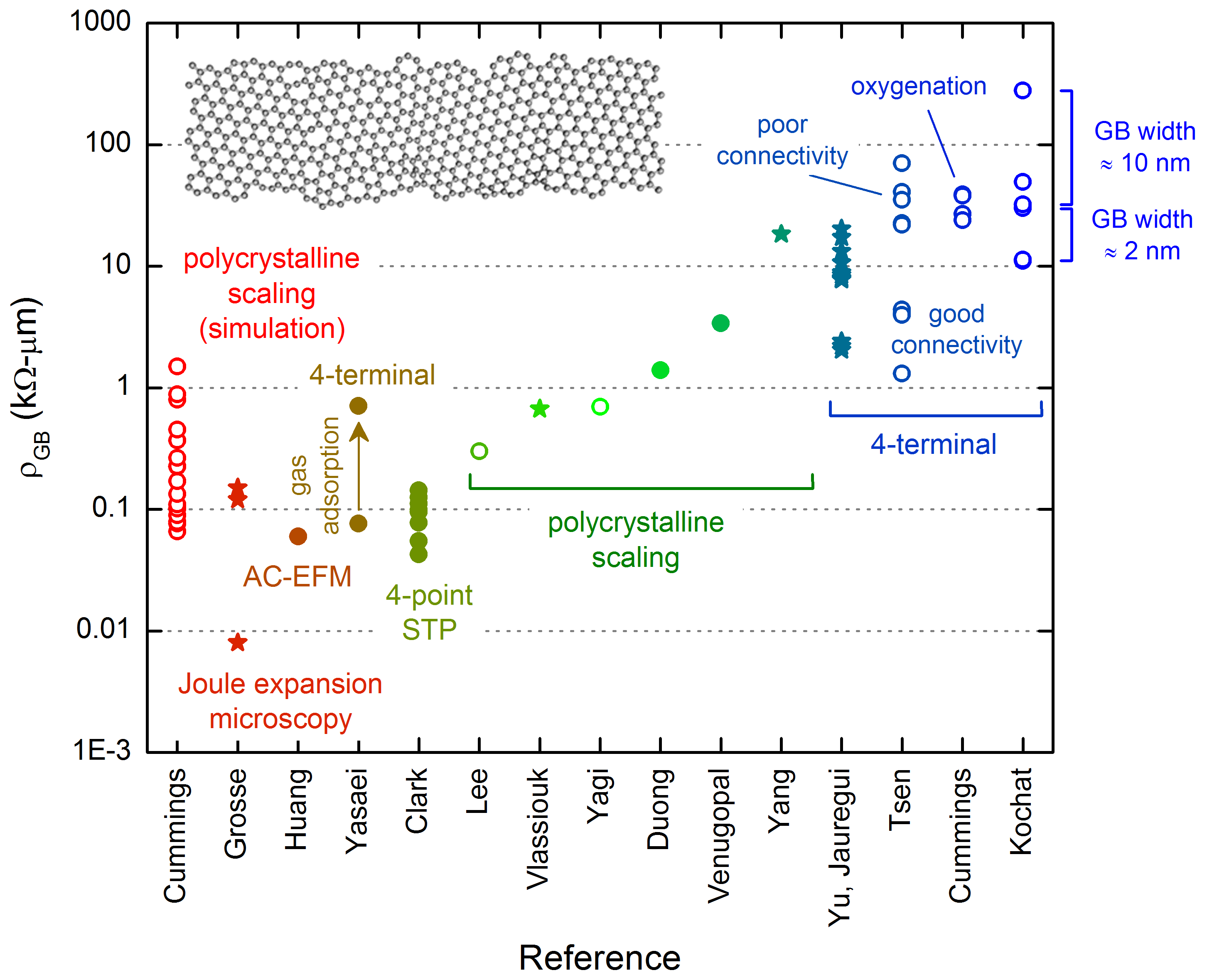}
\caption{Summary of the values of grain boundary resistivity
  ($\rho_{\rm GB}$) extracted from the literature \cite{Cummings2014,
    Grosse2014, Huang2011, Yasaei2014, Clark2013, Lee2014,
    Vlassiouk2011, Yagi2013, Duong2012, Venugopal2012, Yang2016,
    Yu2011, Jauregui2011, Tsen2012, Kochat2016}. Each value is labeled
  according to the measurement technique, and additional notes have
  been included where appropriate. Open circles represent measurements
  made at the charge neutrality point, closed circles represent
  measurements made far from the charge neutrality point, and
  $\star$'s are for measurements where the position of the Fermi level
  is unknown. The spread of simulation results is due to the impact of
  chemical functionalization on the value of $\rho_{\rm GB}$.}
\label{fig:pGB}
\end{figure}

A summary of the values of $\rho_{\rm GB}$ that have been reported in
the literature is shown in Fig. \ref{fig:pGB}. Markedly, $\rho_{\rm
  GB}$ spans more than three orders of magnitude, from less than 0.1
up to 100 k$\Omega$-$\mu$m. To account for this spread of values,
there are several factors that should be considered. Factors that
impact the intrinsic value of the GB resistivity include the
structural quality of the GB, the position of the Fermi level, and the
measurement technique. Other factors that can alter the GB resistivity
are related to the cleanliness of the device, such as if the graphene
was protected by a gate dielectric or if an exposed graphene film was
measured in air or in vacuum.

Tsen {\it et al} showed that poorly-connected GBs can significantly
enhance $\rho_{\rm GB}$, in their case by one order of magnitude
\cite{Tsen2012}, and this behavior has also been seen in numerical
simulations \cite{Tuan2013}. Kochat {\it et al} also showed a strong
correlation between the GB quality and its resistivity, where GBs with
a wide region of disorder were more resistive than narrower GBs
\cite{Kochat2016}. Meanwhile, the numerical simulations in
Fig. \ref{fig:pGB} were made with perfectly connected polycrystalline
samples, and thus yielded relatively low values of $\rho_{\rm GB}$
\cite{Cummings2014}.

When determining the GB resistivity, control of the Fermi level is
also important, as four-terminal measurements have shown that the
value of $\rho_{\rm GB}$ can vary by one order of magnitude as a
function of gate voltage, with the maximum occurring at the charge
neutrality point (CNP) \cite{Tsen2012, Kochat2016}. In
Fig. \ref{fig:pGB}, open circles represent measurements with the Fermi
level at the CNP, closed circles are for measurements where the Fermi
level is far from the CNP, and $\star$'s are for measurements where
the Fermi level is unknown. In general, the largest measured values of
$\rho_{\rm GB}$, on the far right of Fig. \ref{fig:pGB}, were all made
at the CNP, while the lowest measured values of $\rho_{\rm GB}$ were
made far from the CNP. Scaling the lowest measured values of
$\rho_{\rm GB}$ by a factor of 10 brings them into the range of the
four-terminal measurements made at the CNP.

The measurement technique also appears to have some impact on the
estimated value of the GB resistivity. In particular, the values of
$\rho_{\rm GB}$ extracted from the scaling law tend to be lower than
those from four-terminal measurements, even at the CNP. When
considering electrical transport through a large polycrystalline
sample, charge will flow through a series of parallel conducting
paths, and the sheet resistance will be determined by the lowest
resistance path. For this reason, the value of $\rho_{\rm GB}$
extracted from the scaling law represents the lower end of the range
of GB resistivities present in the sample, and not the average
value. When considering all three factors together, it appears that
the resistivity of well-connected GBs, measured at the CNP, typically
falls in a range around 1 k$\Omega$-$\mu$m.

While GBs by themselves may not always give rise to significant
scattering, the adsorption and reaction of adsorbates or impurities
can adversely impact the electrical transport. While this may be
detrimental for high-performance electronic devices, it could be
advantageous for gas sensing applications \cite{Salehi2012,
  Yasaei2015}. Four-terminal measurements of highly-resistive GBs
showed that oxygenation can vary $\rho_{\rm GB}$ by a factor of two
\cite{Cummings2014}, while adsorption of dimethyl methylphosphonate
gas molecules increased the resistivity of a low-resistance GB by one
order of magnitude \cite{Yasaei2014}. Numerical calculations also
revealed a strong variation of $\rho_{\rm GB}$ with chemical
functionalization \cite{Cummings2014}, indicated by the spread of
values on the far left of Fig. \ref{fig:pGB}. In contrast to these
results, recent work by Gao {\it et al} demonstrated a room
temperature mobility of nearly 5000 cm$^2$/V-s for polycrystalline
samples with an average grain size on the order of 100 nm
\cite{Gao2016}. These promising results highlight the need to design
experiments that can separate the intrinsic effect of the GBs from
extrinsic factors like surface cleanliness. The dry transfer method
recently reported by Banszerus {\it et al}, which yielded single-grain
CVD graphene with mobilities up to 350,000 cm$^2$/V-s, is one such
approach \cite{Banszerus2015}.

Beyond impeding charge transport, individual graphene GBs can impact
the electrical properties of polycrystalline graphene in other
ways. For example, Joule expansion microscopy measurements have
revealed strong Joule heating localized at graphene GBs, which can be
many times larger than Joule heating in the grains. This has important
implications for the reliability of graphene devices, as localized
device failure could occur without a significant increase in the
average device temperature \cite{Grosse2014}. Other measurements have
shown that electrical noise is greatly enhanced by graphene GBs, which
is detrimental for low-noise devices but may be useful for sensor
applications \cite{Kochat2016}.

\subsection{Global transport properties of polycrystalline graphene}
\begin{figure}[t]
\includegraphics[width=\columnwidth]{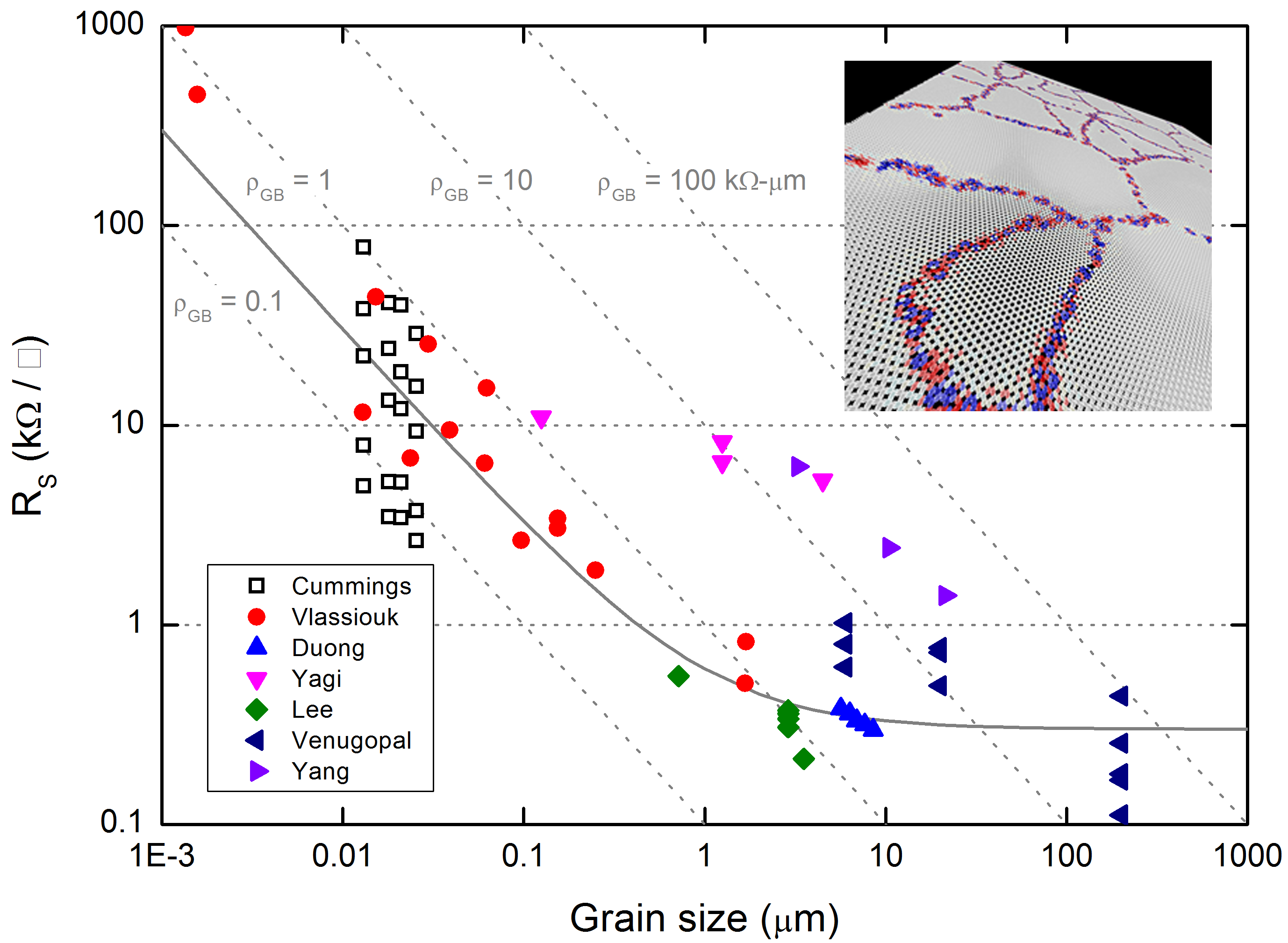}
\caption{Summary of the measured values of polycrystalline graphene
  sheet resistance as a function of average grain size. The solid
  symbols are experimental measurements \cite{Vlassiouk2011,
    Duong2012, Yagi2013, Lee2014, Venugopal2012, Yang2016} and the
  open squares are numerical calculations from
  Ref. \cite{Cummings2014}. The grey solid line shows the behavior of
  the ohmic scaling law, $R_{\rm S} = R_{\rm S}^0 + \rho_{\rm GB} /
  l_{\rm G}$, assuming $R_{\rm S}^0$ = 300 $\Omega / \Box$ and
  $\rho_{\rm GB} = 0.3$ k$\Omega$-$\mu$m.}
\label{fig:RS}
\end{figure}

While knowledge of the resistivity of individual GBs is valuable from
both a fundamental point of view and for particular applications, it
is also important to have a clear picture of the global charge
transport properties of polycrystalline graphene. This is especially
important for large-area devices that utilize this material, such as
flexible transparent electrodes. In polycrystalline graphene, charge
transport is ultimately limited by two sources -- scattering at the
GBs, and scattering within the graphene grains. The competition
between these two sources of resistance is simply captured by the
ohmic scaling law mentioned above, $R_{\rm S} = R_{\rm S}^0 +
\rho_{\rm GB}/l_{\rm G}$. For samples with large grains, the sheet
resistance will be dominated by that of the grains, $R_{\rm S}^0$, and
$R_{\rm S}$ will be independent of the grain size. For small-grained
samples, $R_{\rm S}$ will be dominated by $\rho_{\rm GB}$ and will
scale inversely with the grain size. The crossover from grain- to
GB-dominated behavior occurs at a grain size of $l_{\rm G} \approx
\rho_{\rm GB} / R_{\rm S}^0$. This general behavior is shown by the
solid gray line in Fig. \ref{fig:RS}, where we plot the sheet
resistance as a function of the grain size assuming $R_{\rm S}^0$ =
300 $\Omega / \Box$ and $\rho_{\rm GB} = 0.3$ k$\Omega$-$\mu$m.

In Fig. \ref{fig:RS} we also plot a selection of values of sheet
resistance vs. grain size that have been reported in the experimental
literature. Most measured values follow a general inverse scaling
between $R_{\rm S}$ and $l_{\rm G}$, and the slope of this scaling can
provide information about the relative values of $\rho_{\rm GB}$ and
$R_{\rm S}^0$. For example, the shallow scaling reported by Yagi {\it
  et al} \cite{Yagi2013} suggests a low value of $\rho_{\rm GB}$ = 0.7
k$\Omega$-$\mu$m, but a relatively large grain sheet resistance of
$R_{\rm S}^0$ = 6 k$\Omega / \Box$. However, the measurements of Yang
{\it et al} \cite{Yang2016}, while similar in magnitude, show a much
steeper slope, giving a large value of $\rho_{\rm GB}$ = 18
k$\Omega$-$\mu$m but a smaller $R_{\rm S}^0$ = 600 $\Omega /
\Box$. The measurements of Duong {\it et al} \cite{Duong2012}, Lee
    {\it et al} \cite{Lee2014}, and Venugopal \cite{Venugopal2012} all
    reveal low values of $R_{\rm S}^0$ = 100-300 $\Omega / \Box$,
    while those of Lee {\it et al} have the smallest GB resistivity,
    0.3 k$\Omega$-$\mu$m \cite{Lee2014}. It should be noted that for
    the values reported by Vlassiouk {\it et al}, Raman spectroscopy
    was used to determine the size of defect-free regions, but this
    technique does not distinguish between disorder in the grains and
    disorder arising from the GBs \cite{Vlassiouk2011}. As mentioned
    before, the spread in the numerical calculations of the sheet
    resistance (open squares) is due to varying degrees of chemical
    functionalization applied to the GBs \cite{Cummings2014}.

It is clear that minimizing the impact of GBs is important for
electronic applications of polycrystalline graphene, and today it is
possible to grow single graphene grains with diameters on the order of
centimeters \cite{Li2015, Wu2015, Lin2016}. However, there is a
significant tradeoff between grain size, the required temperature, and
the growth time in the CVD process, and when it comes to industrial
applications these growth conditions should be minimized
\cite{Ryu2014}. By understanding the scaling of the sheet resistance,
one can determine the grain size needed for a particular
application. For example, with a growth process that yields a grain
sheet resistance of $R_{\rm S}^0$ = 100 $\Omega / \Box$ and a GB
resistivity of $\rho_{\rm GB}$ = 1 k$\Omega$-$\mu$m, the grains only
need to be larger than $\rho_{\rm GB} / R_{\rm S}^0$ = 10 $\mu$m for
the contribution of the GBs to no longer matter. Thus, depending on
the requirements, centimeter-sized grains may be unnecessary. This was
recently demonstrated by Samsung and its collaborators, whose rapid
thermal CVD process yielded polycrystalline graphene samples with a
sheet resistance that was constant over grain sizes in the range of
1-10 $\mu$m \cite{Ryu2014}.

In addition to direct charge transport, polycrystalline graphene has
also been studied for its application to quantum Hall metrology. In
single-crystal graphene, clear signatures of the quantum Hall effect
have been measured at room temperature \cite{Novoselov2007}, owing to
the relatively large splitting of the two lowest Landau levels. This
makes graphene quite promising for the establishment of new resistance
standards, as measurements can be made at higher temperatures and
lower magnetic fields than those in traditional two-dimensional
electron gases \cite{Poirier2010, Janssen2015}. However, in
polycrystalline graphene the quantum Hall measurements are less ideal,
with a degradation of the precision of the quantum Hall plateaus and
the development of a finite longitudinal conductivity \cite{Cao2010,
  Shen2011, LaFont2014}. Various numerical simulations have shown that
GBs can short circuit the quantum Hall measurement
\cite{Cummings2014b, LaFont2014, Bergvall2015, Lago2015}, and the
network of GBs provides a path for percolating transport through the
bulk of the material \cite{Cummings2014b}. Very small grains can also
impede the onset of the quantum Hall regime until the magnetic field
becomes large enough such that $l_{\rm B} < l_{\rm G}$, where $l_{\rm
  B} = \sqrt{\hbar / eB}$ is the magnetic length \cite{Cummings2014b,
  Gao2016}. In general, these results indicate that GBs should be
avoided in order to achieve high-precision quantum Hall measurements.

Polycrystalline graphene has also been considered for thermoelectric
applications. The figure of merit of a thermoelectric material is
given by $ZT = S^2 \sigma T / \kappa$, where $T$ is the temperature,
$S$ is the Seebeck coefficient, and $\sigma$ ($\kappa$) is the
electrical (thermal) conductivity. $ZT$ is a measure of the efficiency
of a thermoelectric engine, and its value can be tuned by separately
tuning $S$, $\sigma$, and $\kappa$. Recent theoretical work has
predicted that an individual GB can significantly enhance the $ZT$ of
graphene at the CNP, primarily through an enhancement of the Seebeck
coefficient \cite{Lehmann2015}. However, simulations of large-area
polycrystalline graphene samples are less optimistic, as they show
that $\sigma$ and $\kappa$ scale similarly with the grain size
\cite{Tuan2013, Mortazavi2014, Hahn2016_1, Hahn2016_2}, while the
Seebeck coefficient is reduced compared to pristine graphene
\cite{Woessner2016}. Together, these results suggest an overall
reduction of $ZT$ compared to single-grain graphene.

Finally, it is important to note that beyond the microstructure, the
process of device fabrication can also have a strong impact on
electrical transport. Exfoliated and CVD graphene are usually
transferred onto a dielectric substrate using scotch tape or organic
compounds such as PMMA, resulting in physisorbed or chemisorbed
residues on the graphene surface \cite{Pirkle2011}. If the graphene
sheet contains point defects, dislocations, grain boundaries, or
wrinkles, the residues can attach more easily \cite{Seifert2015}. In
fact, during the initial stages of graphene device fabrication,
dielectric films were usually difficult to deposit on unfunctionalized
or "clean" graphene surfaces, and several techniques were introduced
to overcome this problem \cite{Lee2008, Kim2009, Farmer2009}. The
ability to deposit the dielectric uniformly on graphene protected the
films from ambient adsorbates, which can have a significant effect on
the transport properties \cite{Chan2012}, but these processes also
introduced organics, ozone, metals, or metal oxides that tended to
degrade the mobility in comparison to devices without a top gate
\cite{Fallahazad2010, Fallahazad2012}. Therefore, when analyzing
graphene mobility data it is critically important to know the quality
of the dielectric substrate, the crystallinity of the graphene, and
the gate dielectric. This was clearly demonstrated by the use of
hexagonal boron nitride (h-BN) as a substrate material and gate
dielectric \cite{Dean2010}, which can lead to very high mobilities in
both single-crystal and bi-layer graphene \cite{Hao2013, Hao2016}. In
cases where the surface of graphene is not exposed to any organic
compounds during the transfer process, the mobility is also greatly
improved \cite{Banszerus2016} although in this case perhaps the
benefit is also associated with low strain in the graphene films
transferred using h-BN \cite{Neumann2015}.

\section{Mechanical properties of polycrystalline graphene}
Graphene is often noted for its remarkable mechanical properties,
including its large elastic modulus and high fracture strength. For a
low-dimensional material such as graphene it is natural to expect
grain boundaries to greatly alter elastic constants and the fracture
strength. Several experimental, as well as theoretical works, suggest
that individual grain boundaries can possess mechanical properties
close to those of pristine
graphene~\cite{Lee2013,Yazyev2014,Ophus2015}. However, other works,
with a more global view on polycrystalline graphene, indicate that
these properties can strongly depend on grain size $l_{\rm G}$ and
system size $L$. In particular, the ratio $l_{\rm G}/L$ may play an
important role, and few works have so far been able to access the
limit of $l_{\rm G}/L\ll1$, which is of importance for many
large-scale applications.

\subsection{Influence on elasticity}
The elastic properties of graphene are commonly studied experimentally
by means of nanoindentation experiments~\cite{Lee2008_2}. For
numerical studies, molecular dynamics (MD) is typically required as
system sizes are in general too large for density functional theory
(DFT) to be practical. In a nanoindentation measurement, the graphene
devices have the form of a micron-sized suspended circular drum. By
pressing down with an atomic force microscope (AFM) while recording
applied force and tip deflection, the elastic properties are
deduced. While experimental devices have micron linear dimension,
numerical simulations are limited to much smaller systems ($\lesssim
100$~nm). In MD studies, the mechanical properties are probed by
recording the stress as the system is uni- or bi-axially
strained. This should be contrasted with the nanoindentation
experiments where the strain is typically non-uniform even for a
pristine sheet and where the details of the AFM tip shape may
influence the results. For pristine graphene, both numerics (MD/DFT)
and indentation experiments show good agreement, yielding a Young's
modulus either quoted as a 2D entity $E_{\rm 2D}\approx 340$~N/m or as
the equivalent 3D modulus $E_{\rm 3D}\approx
1$~TPa~\cite{Kudin2001,Lee2008_2}. The two are related through the
interlayer spacing of graphite ($3.35$~{\AA}).

For polycrystalline sheets, however, there is a larger discrepancy
between reported values. On the experimental side, one complication
arises from the different methods of fabrication and transfer. Not
only does this yield samples with different quality of grain
boundaries and different amounts of pre-stress, but also the amount of
wrinkling differs between samples. Wrinkling or buckling can be either
the result of the fabrication-transfer process or due to the
boundaries themselves, which alleviate the extra strain incurred by
the defects. This causes a softening of the Young's modulus for small
strains. On the theoretical side, the choice of interaction potential
and grain boundary geometry can greatly influence the results.

Theoretical studies of the simplest geometry, graphene sheets
containing a single straight grain boundary~
\cite{Grantab2010,Wei2012,Liu2012}, found values around $E_{\rm
  3D}\sim 800$~GPa. This was confirmed by nanoindentation measurements
using an AFM tip to press down on polycrystalline suspended CVD
graphene drums which showed, within statistical error, a value of
$E_{\rm 3D}\sim 1$~TPa~\cite{Lee2013}. Although the spread in measured
elastic constants was larger for polycrystalline graphene than for
pristine graphene, the average value was still in agreement with that
of pristine graphene. In the experiments by Lee {\it et
  al}~\cite{Lee2013}, the diameter of the suspended graphene drums was
similiar to the grain size.

\begin{figure}[t]
\begin{center}
\includegraphics[width=.9\columnwidth]{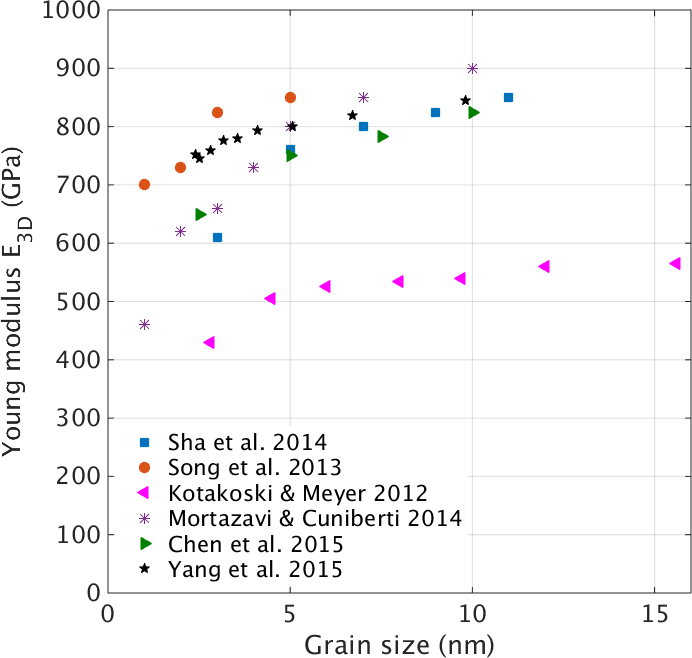}
\end{center}
\caption{Young's modulus for polycrystalline graphene as function of
  linear grain size as obtained from MD
  simulations~\AI{\cite{Kotakoski2012,Song2013,Sha2014,Chen2015,Yang2015,
      Mortazavi2014_2}}. While there is a large discrepancy in
  absolute numbers, all studies show a trend of increasing stiffness
  with grain size.}
\label{fig:EGB}
\end{figure}

Both the theory~\cite{Grantab2010,Wei2012,Liu2012} and
experiments~\cite{Lee2013} essentially studied local grain boundary
properties, i.e they were in the regime $l_{\rm G}/L\lesssim
1$. However, those findings are in contrast to theory taking into
account more natural grain boundary geometries. Such simulations
typically result in lower values for $E_{\rm
  3D}$~\AI{\cite{Kotakoski2012,Song2013,Sha2014,Chen2015,Yang2015,Mortazavi2014_2}}. In
particular, Kotakoski and Meyer~\cite{Kotakoski2012}, who performed
large-scale MD simulations on polycrystalline graphene, found
significantly lower values of the Young's modulus $E_{\rm 3D}\sim
600$~GPa. While smaller values were obtained experimentally using
indentation measurements~\cite{Huang2011,Ruiz-Vargas2011} with values
$E_{\rm 3D}\sim 150$~GPa, these were attributed the to the out-of
plane buckling associated with the grain boundaries. Finally, not only
grain boundaries may affect the elastic properties. As highlighted by
nanoindentation measurements~\cite{Zandiatashbar2014,Lopez-Polin2015}
on graphene with controlled creation of defects, the type of defects
and their concentration can also impact the results.

As in 3D macroscopic materials, the characteristic grain size $l_{\rm
  G}$ plays an important role. From MD simulations, a clear grain-size
dependence of the elastic modulus has been found, with $E_{\rm 3D}$
increasing upon increasing the grain
size~\AI{\cite{Kotakoski2012,Song2013,Sha2014,Chen2015,Yang2015,Mortazavi2014_2}}
(see Fig.~\ref{fig:EGB}). In order to have a well-defined value of an
elastic property, it is desirable that it be measured for a system
size $L\gg l_{\rm G}$. For practical reasons, simulating structures
with linear dimensions $L$ much larger than the characteristic grain
sizes has remained prohibitive except for very small grains ($\sim
1$~nm), and most studies have not studied the scaling with $l_{\rm
  G}/L$. Also, nanoindentation studies are performed on
micrometer-sized systems comparable to the grain sizes ($L\sim l_{\rm
  G}\sim 1$~$\mu$m), and reporting the ensemble-averaged values over
many devices. There is, however, currently no experimental work
covering the case $L\gg l_{\rm G} \sim 1$~$\mu$m for the Young's
modulus.

\begin{figure}[t]
\begin{center}
\includegraphics[width=.86\columnwidth]{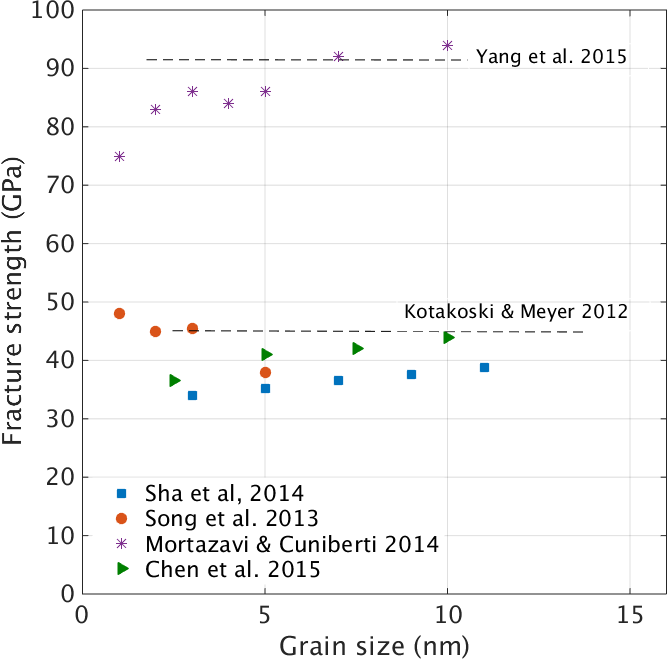}
\end{center}
\caption{Fracture strength obtained by MD as a function of grain size
  from
  references~\AI{\cite{Sha2014,Song2013,Mortazavi2014_2,Chen2015}}. While
  three of the studies~\AI{\cite{Mortazavi2014_2,Sha2014,Chen2015}}
  show increasing fracture strength with increasing grain size, the
  results of Song {\it et al}~\cite{Song2013} show the opposite
  trend. The grain-size-independent results by Kotakoski and
  Meyer~\cite{Kotakoski2012} and Yang {\it et al}~\cite{Yang2015} are
  shown as dashed lines.}
\label{fig:fracture}
\end{figure}

\subsection{Fracture strength}
Monocrystalline graphene is predicted to have an extraordinary
fracture strength of 110 -
130~GPa~\cite{Liu2007,Wei2012,Lee2013,Rajasekaran2015}. In presence of
grain boundaries it is natural to expect that this strength is
diminished. However, using MD, it was shown~\cite{Grantab2010} that in
contrast to this expectation, higher-angle tilt grain boundaries with
a higher density of defects (heptagon-pentagon pairs) can be as strong
as mono-crystalline graphene, while low-angle boundaries with lower
defect concentrations showed a reduction in fracture strength. This
behaviour derives from the dipolar stress profile around individual
pentagon-heptagon pairs surrounded by hexagons, with tensile stress on
the heptagon side and compressive stress on the pentagon side (see
Fig.~\ref{fig:GB}c). With an increased density of aligned
pentagon-heptagon pairs, stress cancellation from adjacent dipoles
results in reduced built-in stress. These results were confirmed by
studies~\cite{Wei2012,Liu2012,Zhang2012} which further found that the
detailed arrangement of defects in the GB played an important role. In
particular, Wei {\it et al}~\cite{Wei2012} found a non-monotonic
behaviour of fracture strength with tilt angle for armchair grain
boundary configurations, which was also corroborated by a continuum
theory in excellent agreement with simulations. Subsequent MD studies
have revealed similar
results~\cite{Yi2013,Jhon2013,Wu2013,Han2014,Yang2014}.  While most studies used
straight boundaries, Rasool {\it et al}~\cite{Rasool2014} and Ophus
{\it et al}~\cite{Ophus2015} showed that more complex boundary
structures also typically exhibited a high fracture strength $\sim
100$~GPa. Using MD it has further been shown that the fracture
strength of tilt grain boundaries decreases significantly with
increasing temperature~\cite{Yi2013, Zhang2013, Yang2014, Chen2015}.

Experimental determination of fracture strength has been carried out
by means of nanoindentation measurements. Early
results~\cite{Huang2011, Ruiz-Vargas2011} revealed a significantly
decreased fracture strength $\sim 35$~GPa, which was attributed to
vacancies and shear stress along the boundaries. However, by using
different post-processing methods, it was shown, in accordance with
the theoretical predictions, that well-stitched boundaries could have
nearly the same fracture strength ($\approx 100$~GPa) as pristine
graphene~\cite{Lee2013}. Rasool {\it et al}~\cite{Rasool2013}
characterised the grain boundary angles in their measurements,
confirming the angle dependence of the fracture strength, and finding
high values close to 100~GPa for large angle boundaries. To explain
the observed spread in fracture strength in a nanoindentation
measurement, Sha {\it et al}~\cite{Sha2014_2} made explicit MD
simulations of nanoindentation measurements of polycrystalline
graphene, revealing that the relative positioning of the tip and the
grain boundaries could yield vastly different fracture strength and
different failure paths.

The dependence of fracture strength on tilt angle was predicted by
simulations of bi-crystal systems with a single boundary, but more
realistic fracture strength simulations of polycrystalline graphene
have shown a significant reduction in fracture
strength~\cite{Kotakoski2012} and dependence on grain
size~\cite{Sha2014, Mortazavi2014_2, Song2013, Cao2013, Chen2015}. In
these simulations, failure is typically initiated at the boundary, or
more commonly at meeting points between three different grains. The
dependence of fracture strength on grain size for the studies in
Refs.~\cite{Sha2014,Mortazavi2014_2,Song2013,Chen2015} are shown in
Fig.~\ref{fig:fracture}. Notably, three
studies~\cite{Sha2014,Mortazavi2014_2,Chen2015} report increasing
strength with increasing grain size, while Song {\it et
  al}~\cite{Song2013} find the opposite trend. The fracture strengths
in the study by Kotakoski and Meyer~\cite{Kotakoski2012} and the paper
of Yang {\it et al}~\cite{Yang2015}, however, showed no significant
dependence on grain size.

Nanoindentation measurements by Suk {\it et al}~\cite{Suk2015} give
support to a decreased fracture strength of polycrystalline graphene,
with the fracture strength being smaller for smaller grains. For the
comparison between theory and experiment, it is again worth pointing
out that most numerical studies have $L\gtrsim l_{\rm G}\sim 1$~nm,
while experimental works are in the regime $L\sim l_{\rm G}\sim
1$~$\mu$m. In a recent study by Shekhawat and
Ritchie~\cite{Shekhawat2016}, a weakest-link argument is used together
with MD simulations of $\sim2\times10^4$ grain boundary configurations
to obtain a functional form for the fracture strength. They find a
scaling for the fracture strength $\sigma$, $\sigma-\sigma_0\propto
\nu(l_{\rm G}/L)^{2/m}$. Here $\sigma_0=19.5$~GPa, $\nu=53.2$~Gpa and
$m\approx 10.1$, showing that the fracture strength of large sheets
may be significantly lower than previously predicted.

\section{Thermal transport properties}
The experimental determination of the lattice thermal conductivity
$\kappa$ in graphene remains a matter of debate, since heat transport
in such an atomic sheet is sensitive to many details, including the
sample dimension and boundaries, point-like or extended defects, and
whether the sample is suspended or supported \cite{Balandin2011}. The
estimation of $\kappa$ is also affected by the experimental protocol,
which includes both steady--state measurements \cite{Balandin2008,
  Ghosh2008, Lee2011, Jang2013} and measurements in the transient
regime where the system evolves from an initial condition of thermal
nonequilibrium \cite{Jang2010,Cabrera2015,Guo2008a,Guo2008b}. This
wide range of scenarios is reflected in reported experimental values
of $\kappa$ in the interval of $1000-5000\ {\rm WK^{-1}m^{-1}}$.

The theoretical prediction of $\kappa$ is also controversial, since a
direct comparison among unlike results is made difficult by different
adopted simulation protocols. This includes approximate or exact
solutions of the Boltzmann transport equation \cite{Lindsay2010a,
  Lindsay2010b,Singh2011,Lindsay2014,Cepellotti2014} (BTE) and
molecular dynamics (MD) simulations in different flavors, such as
equilibrium \cite{Pereira2013,Fan2015} (EMD), nonequilibrium
\cite{Xu2014} (NEMD), or approach-to-equilibrium \cite{Barbarino2015}
(AEMD) formulations. Both full ab initio \cite{Lindsay2014,
  Cepellotti2014, Fugallo2014} calculations and MD simulations based
on empirical potentials have been published, the latter being mostly
based on the Tersoff force field \cite{Lindsay2010c} or the reactive
empirical bond order potential (REBO) \cite{Brenner2002}. In summary,
theoretical results for $\kappa$ at room temperature vary in the range
$1000\ {\rm WK^{-1}m^{-1}}\le \kappa \le 8000\ {\rm WK^{-1}m^{-1}}$
\cite{Nika2012}.

Finally, we remark that the direct comparison between theoretical
predictions and experimental results is often questionable, since most
calculations are carried out in idealized situations missing many of
the structural details ruling over the experiments. As a matter of
fact, real graphene samples, fabricated either by epitaxial film
growth \cite{Berger2006, Hass2008} or CVD \cite{Li2009, Ding2011,
  Su2011}, are hardly pristine due to limitations in the growth
process and because of substrates and, therefore, they are
characterized by defects limiting the size of pristine crystalline
domains. Their multi-grain structure, with dimensions down to the
micro- and nanoscale, is likely an important pre-existing cause for
the wide range of $\kappa$ values reported above. However, the
systematic investigation of the effect of size, shape, and
distribution of grains on thermal transport properties is still
ongoing \AI{\cite{Mortazavi2014, Liu2014, Hahn2016_1, Hahn2016_2,
    Bagri2011, Wu2014, Wang2014}}.

\subsection{Thermal resistance of a single grain boundary}
As described in section~\ref{sec:GBconf}, a single grain boundary can
be created in the honeycomb lattice by tilting two crystalline
graphene sheets of the same length by a certain angle $\alpha$ (see
Fig.~\ref{fig:GB}b). Although such simple geometries correspond to an
idealized configuration of a perfectly straight, periodic and isolated
grain boundary, they nevertheless represent a paradigmatic situation
for investigating the role of GBs in thermal transport.

\begin{figure}[t]
\centering
\includegraphics[width=\linewidth]{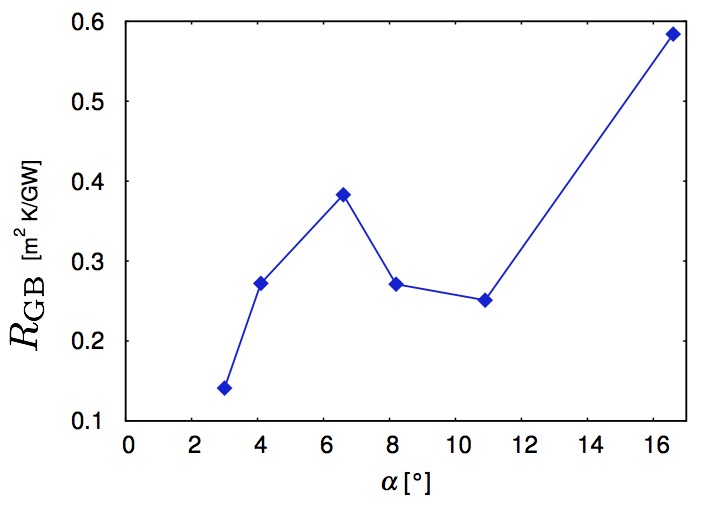}
\caption{Predicted grain boundary thermal resistance $R_{\rm GB}$ as
  function of the tilt angle $\alpha$ defining the structure of the
  grain boundary (see Fig.~\ref{fig:GB}b).\label{fig:LCfig2}}
\end{figure}

Both the strain-induced corrugation and the lattice misorientation
between the two neighboring grains deeply affect the heat transport
along the direction normal to the grain boundary, resulting in an
effective GB thermal resistance $R_{\rm GB}$. This can be described as
a series of resistances \cite{Mortazavi2014} according to
\begin{equation}\label{eq:Rthermal1}
R_{\rm tot} = \frac{L_{\rm tot}}{\kappa_{\rm tot}}=\frac{L_{\rm left}}{\kappa_{\rm left}} + R_{\rm GB} + \frac{L_{\rm right}}{\kappa_{\rm right}},
\end{equation}
where $R_{\rm tot}$, $L_{\rm tot}$, and $\kappa_{\rm tot}$ are the
thermal resistance, length, and thermal conductivity of the total
sample, while $L_{\rm left, right}$ and $\kappa_{\rm left, right}$
represent the length and thermal conductivity of the left and right
crystallites, respectively (see Fig.~\ref{fig:GB}b).  By selecting a
suitable simulation cell with $L_{\rm right}=L_{\rm left}=L_{\rm
  tot}/2$ and $\kappa_{\rm left}=\kappa_{\rm right}=\kappa_{\rm sg}$
(here $\kappa_{\rm sg}$ represents the thermal conductivity of a
single-grain sample of pristine graphene), Eq. \ref{eq:Rthermal1}
allows one to predict the GB thermal resistance as
\begin{equation}\label{eq:Rthermal2}
R_{\rm GB} = \frac{L_{\rm tot}}{2} \left ( \frac{1}{\kappa_{\rm tot}} -  \frac{1}{\kappa_{\rm sg}} \right ).
\end{equation}
All quantities appearing in Eq. \ref{eq:Rthermal2} are
straightforwardly calculated by an AEMD simulation
\cite{Barbarino2015,Hahn2016_1}, which provides the value of
room-temperature $R_{\rm GB}$ as a function of the tilt angle
$\alpha$. \AI{As shown in
  Fig.~\ref{fig:LCfig2}}, the trend is not monotonic, which is likely
due to a complex interplay among buckling effects, the occurrence of
coordination defects, and bond-network reconstruction. This
non-monotonic behavior has not been seen experimentally, but the
values of $R_{\rm GB}$ are of the same order of the available
experimental data reported in Ref.~\cite{Yasaei2015}. Furthermore, the
same experiments confirm the general trend of increasing thermal
boundary resistance with increasing tilt angle. \AI{This is consistent
  with previous findings reporting an increasing effective thermal
  conductivity with decreasing mismatch angle~\cite{Wang2014}.} We
remark that increased thermal boundary resistance in multi-grain
graphene has been attributed to larger out-of-plane buckling
\cite{Hahn2016_2,Helgee2014}. This, however, is not valid for the
single isolated GB; the highest thermal boundary resistance occurs for
the largest tilt angle where the lowest out-of-plane buckling is
observed, as shown in Fig.~\ref{fig:GB}d.

\subsection{Thermal conductivity of nanoscrystalline graphene}
\begin{figure}[t]
\centering
\includegraphics[width=\linewidth]{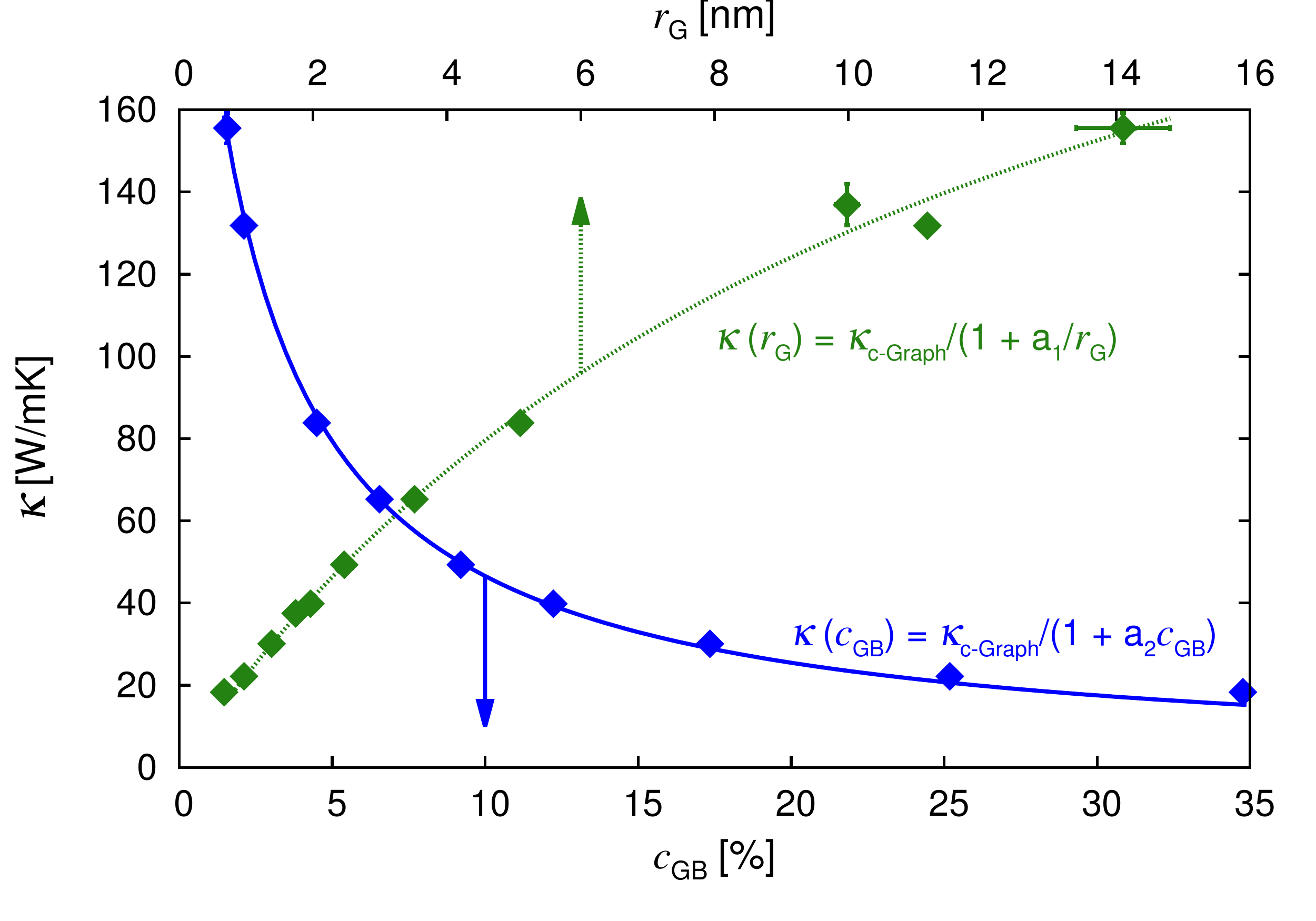}
\caption{Predicted room-temperature thermal conductivity $\kappa$ of
  nanocrystalline graphene as a function of the concentration of atoms
  in the grain boundaries $c_{\rm GB}$ (blue solid line), and of the
  average radius of gyration of the graphene grains $r_{\rm G}$ (green
  dotted line). Simulations correspond to samples as long as 200
  nm. \label{fig:LCfig3}}
\end{figure}

The thermal conductivity in model nanocrystalline graphene samples has
been calculated in \cite{Hahn2016_1, Hahn2016_2} and the main results
are summarized in Fig.~\ref{fig:LCfig3}, where \AI{the}
room-temperature thermal conductivity $\kappa$ is reported as a
function of the average radius of gyration $r_{\rm G}$ of the grains
(measuring the average granulometry of the nanocrystalline sample) or,
equivalently, as a function of the concentration $c_{\rm GB}$ of atoms
in the grain boundary (defined as belonging to non-hexagonal rings in
the graphene lattice). As expected, samples with a higher density of
grain boundaries (or, equivalently, a higher concentration of GB
atoms) correspond to smaller thermal conductivity values, consistent
with the analysis presented in the previous Section. In all cases,
$\kappa$ is very much reduced with respect to the value 262
W/(m$\,$K), found for a single-grain graphene sample with same
length. \AI{This is in good agreement with Ref.~\cite{Wu2014} where a
  reduction of thermal conductivity to about 20 \% of the value in
  pristine graphene is observed for an average grain size of
  $\sim2.5$~nm.} This highlights the major role played by GBs in
scattering phonons, whose mean free path (MFP) is comparable with the
average GB spacing. Based on the accumulation function of the thermal
conductivity, the average phonon MFP has been estimated to vary from
451 nm for pristine samples to $\sim$ 30 nm for nanocrystalline ones
\cite{Hahn2016_1,Hahn2016_2}.

An inverse rational function can be used to describe the scaling of
$\kappa$ with $r_{\rm GB}$ and $c_{\rm GB}$, as shown in
Fig.~\ref{fig:LCfig3} (where $\kappa_{\rm c-Graph}$ is the thermal
conductivity of the crystalline grains). A similar argument has been
used previously in Ref. \cite{Mortazavi2014}, where the thermal
conductivity of polycrystalline graphene has been calculated by EMD
simulations. The good agreement between the calculated data and
interpolation functions indicates that the effective thermal
conductivity of nanostructured graphene can be estimated considering
grain boundaries and grains as a connection of resistances in
series. Given this, the average GB resistance extracted from
Fig.~\ref{fig:LCfig3} is $\sim$0.1 m$^2\,$K/GW. This is in reasonably
good agreement with \AI{the predictions reported in
  Ref.~\cite{Bagri2011} as well as with} the results of the previous
Section if one considers that the value extracted from the grain size
scaling represents the lower end of the GB resistances in the
sample. Further details are found in Ref. \cite{Hahn2016_1}. In this
scheme, phonon scattering by the GBs is mimicked as an effective
interface thermal resistance, while the thermal conductivity of
crystalline domains is described by the thermal conductivity of the
crystal without grain boundaries.

\section{Summary}
The scaling analysis of the physical properties of polycrystalline
graphene with average grain size has revealed several fundamental
features. First, while charge and thermal transport are generally
found to scale linearly or sub-linearly with increasing grain size,
mechanical properties such as Young's modulus or fracture strength are
less dependent on such parameters. Second, electrical measurements
have found that the resistivity of individual grain boundaries can
vary by up to four orders of magnitude, depending on the quality of
the GB and the measurement conditions. However, for large-area
polycrystalline graphene the average resistivity of the GBs shows much
less variation from sample to sample. Concerning thermal transport,
GBs are very efficient phonon scatterers that dramatically reduce the
thermal conductivity with respect to pristine graphene. Furthermore,
the overall transport properties can be effectively modelled by
looking at a nanocrystalline sample as a series of thermal resistances
attributed to both grains and GBs. In particular, the predicted GB
thermal resistance shows a non-monotonic variation with the tilt
angle, arising from the combined effect of interface buckling,
coordination defects, and bond reconstruction.

The overall set of information reported here gives a comprehensive
picture of the current understanding of the relationship between
polycrystalline morphology and the main physical properties of
large-area CVD graphene. Ultimately, this can help to assess the
usefulness of this material for a variety of applications, from
wearable flexible electronics to biosensors to spintronic
devices. Additionally, this information can help to guide the
production methods and conditions necessary to achieve the material
characteristics desired for a particular application. \AI{This issue
  is common to other 2D materials such as MoS2 where, by combining
  two-laser Raman thermometry with finite element simulations, it has
  recently been shown that heat conduction can be effectively
  engineered by controlling the nanoscale grain
  size~\cite{Sledzinska2016}.} For sensing applications smaller grains
may be desired, as point defects or grain boundaries could improve the
device performance, while for purely electronic applications larger
grains would be ideal. Point defects and grain boundaries could also
act to tune the interaction with underlying substrates, or as
disruptive defects to dictate mean free paths or spin lifetimes
\cite{Roche2015}.

\section*{Acknowledgements}
AWC, JMK, and SR acknowledge support from the European Union Seventh
Framework Programme under grant agreement 604391 Graphene Flagship.
LC, AWC and SR acknowledge support from the Severo Ochoa Program
(MINECO, Grant SEV-2013-0295). AWC and SR acknowledge support from the
Spanish Ministry of Economy and Competitiveness (MAT2012-33911), and
Secretar\'{i}a de Universidades e Investigaci\'{o}n del Departamento
de Econom\'{i}a y Conocimiento de la Generalidad de Catalu{\~{n}}a. LC
acknowledges financial support by the Spanish MINECO under grants
no. FEDER-FIS2012-37549-C05-02, FEDER-MAT2013-40581-P, TEC2012-31330
and TEC2015-67462-C2-1-R, the Generalitat de Catalunya under grants
no. 2014 SGR 301 and 2014 SGR 384, and the Spanish MINECO through the
Severo Ochoa Centres of Excellence Program under SEV-2015-0496. AI
acknowledges support from the Swedish Research Council (VR).
\section*{References} 
\bibliography{GBRev}

\end{document}